\documentclass[osajnl,twocolumn,showpacs, superscriptaddress,10pt]{revtex4-1}
\usepackage{graphicx}%
\usepackage{dcolumn}%
\usepackage{bm}%
\usepackage{amsmath}
\usepackage{siunitx}

\begin{document}
\title{Towards an optical far-field measurement of higher-order multipole contributions to the scattering response of nanoparticles}

\author{T. Bauer}
\affiliation{Max Planck Institute for the Science of Light, Guenther-Scharowsky-Str. 1/Bldg. 24, 91058 Erlangen, Germany}
\affiliation{Institute of Optics, Information and Photonics, University Erlangen-Nuremberg, Staudtstr. 7/B2, 91058 Erlangen, Germany}

\author{S. Orlov}
\affiliation{Max Planck Institute for the Science of Light, Guenther-Scharowsky-Str. 1/Bldg. 24, 91058 Erlangen, Germany}
\affiliation{Institute of Optics, Information and Photonics, University Erlangen-Nuremberg, Staudtstr. 7/B2, 91058 Erlangen, Germany}

\author{G. Leuchs}
\affiliation{Max Planck Institute for the Science of Light, Guenther-Scharowsky-Str. 1/Bldg. 24, 91058 Erlangen, Germany}
\affiliation{Institute of Optics, Information and Photonics, University Erlangen-Nuremberg, Staudtstr. 7/B2, 91058 Erlangen, Germany}
\affiliation{Department of Physics, University of Ottawa, 25 Templeton, Ottawa, Ontario K1N 6N5, Canada}

\author{P. Banzer}
\email{peter.banzer@mpl.mpg.de}
\affiliation{Max Planck Institute for the Science of Light, Guenther-Scharowsky-Str. 1/Bldg. 24, 91058 Erlangen, Germany}
\affiliation{Institute of Optics, Information and Photonics, University Erlangen-Nuremberg, Staudtstr. 7/B2, 91058 Erlangen, Germany}
\affiliation{Department of Physics, University of Ottawa, 25 Templeton, Ottawa, Ontario K1N 6N5, Canada}

\begin{abstract}
We experimentally show an all-optical multipolar decomposition of the lowest-order Eigenmodes of a single gold nanoprism using azimuthally and radially polarized cylindrical vector beams. By scanning the particle through these tailored field distributions, the multipolar character of the Eigenmodes gets encoded into 2D-scanning intensity maps even for higher-order contributions to the Eigenmode that are too weak to be discerned in the direct far-field scattering response. This method enables a detailed optical mode analysis of individual nanoparticles.
\end{abstract}

\maketitle

Geometrically tailored nanostructures can be utilized to create strong local near-field enhancement for light harvesting \cite{Aubry2010,Atwater2010} or higher-harmonic generation \cite{Kim2008,Chen2013}. Furthermore, optimized directive emission of nanoparticles or antennas can be achieved by adapting their amplitude and phase response \cite{Fu2013,Vercruysse2013,Hancu2013,Coenen2014,Neugebauer2014}. The underlying scattering processes are hereby governed by the Eigenmodes of the employed particle at the desired wavelength of operation. A typical example of such a metallic particle with a tailored optical response is a nanoprism. Its  supported electromagnetic multipolar mode structure was shown on the single particle level via near-field measurements \cite{Rang2008} and electron energy loss spectroscopy (EELS) \cite{Nelayah2007}. It is used as a fundamental building block in many areas of nano-optics, plasmonics and sensing \cite{Kim2008,Kinkhabwala2009,Wang2014,Sherry2006} since its spectral response can be tuned easily by changing simple geometric parameters such as the thickness or base-length of the particle \cite{Mock2002,Kelly2003,Nelayah2010}.

\begin{figure}[btp]
\includegraphics[width=\columnwidth]{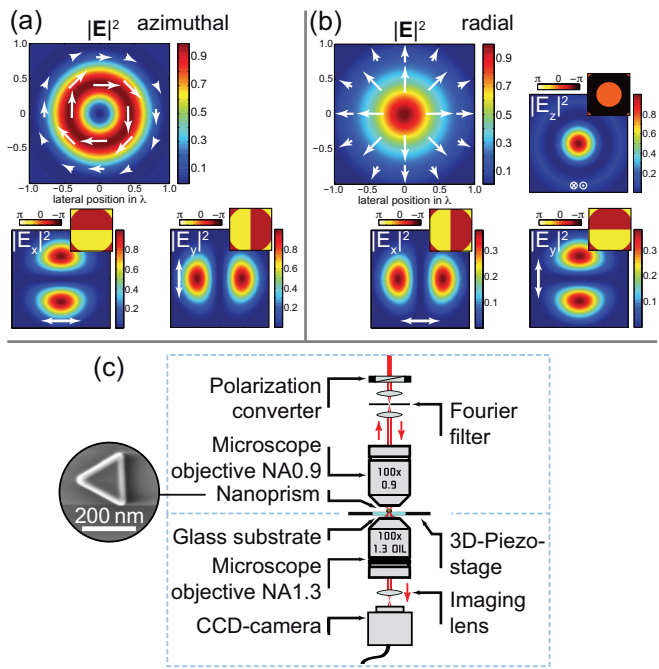}%
\caption{Electric energy density distribution in the focal plane of a tightly focused (a) azimuthally and (b) radially polarized beam and its constituting components (normalized to the maximum total electric energy density) and phases. The white arrows represent a snapshot of the local field direction projected onto the focal plane. The initial beam waist is $w_0 / f = 0.75$, the focusing numerical aperture is 0.9. (c) Sketch of the experimental setup to measure 2D-scanning maps of single metallic nanoprisms.}%
\label{fig:setup}%
\end{figure}

For the investigation of the optical properties of such an individual nanostructure, an all-optical and extraordinarily versatile approach has been presented in literature. This approach relies on the utilization of spatially tailored electromagnetic field distributions at the nanoscale \cite{Banzer2010,Kindler2007,Sancho-Parramon2012}, such as tightly focused cylindrical vector beams (CVB; azimuthally and radially polarized \cite{Quabis2000,Youngworth2000,Dorn2003}, see Fig.\,\ref{fig:setup}(a) and (b)) in the focal plane of a focusing system to study the modal response of individual nanostructures. It was also shown that as a first step, the symmetry and geometry of a single metallic nanoprism can be retrieved from the scattering map recorded while scanning such a prism through the focus of tightly focused CVBs \cite{Zuchner2008}. Each pixel of the two-dimensional scan image contains information about the power collected in transmission or reflection for each positon of the particle scanned through the focal plane. This is possible because the triangular symmetry of the particle of a size comparable to the focal spot gets encoded into the scanning map. For particle dimensions significantly smaller than the optical wavelength, at least asymmetric deformations of the focal spot were experimentally observed in Ref. \onlinecite{Zuchner2008}, allowing for an estimation of the particle's shape and orientation. 
While this trigonal shape of the scattering map seems to be an intuitive match with the geometry of the particle for particle dimensions comparable to the size of the focal spot, it does not correspond to the axial symmetry of the probing field when assuming only local dipolar field interactions for sub-wavelength particles of triangular shape. Within the dipolar approximation, the scattering response does not depend on the electric field orientation relative to the prism. Thus, the scan image should show axial symmetry. To observe the aforementioned triangular pattern of the scan image, the nanoprism has to be sensitive not only to the local electric field (electric dipole response), but also the spatial field distribution due to its chosen lateral dimensions. This is in accordance with reported near-field investigations \cite{Rang2008}. The trifold symmetry of the scattering map is thus a consequence of higher-order multipole contributions to the Eigenmode of the particle for different particle positions.

Here, we want to theoretically and experimentally verify the above-mentioned assumption and show how to all-optically probe this multipolar structure of the supported low-order Eigenmodes in a single sub-wavelength metallic nanoprism from the far-field. By scanning the particle through the focal field distribution of tightly focused CVBs, the resulting scattering scanning map grants access to the actual multipolar response of the investigated nanoprism. In our study, the exemplarily chosen particle is by at least a factor of 3 smaller than the wavelength and the focal spot, hence providing the possibility to access different coupling scenarios for different particle positions. 

For a detailed theoretical investigation of the multipolar Eigenmodes of a single nanoprism in the far-field, the electric field scattered off the nanoprism is decomposed into vector spherical harmonics (VSHs):
\begin{align}
\textbf{E}_\text{sca}(\textbf{r}) = \sum_{n=1}^\infty \sum_{m=-n}^n a_{mn} \textbf{N}_{mn}(\textbf{r}) + b_{mn} \textbf{M}_{mn}(\textbf{r}),
\end{align}
where $\textbf{N}_{mn}$ and $\textbf{M}_{mn}$ represent the electric and magnetic multipole components \cite{Tsang} and $a_{mn}, b_{mn}$ are the corresponding expansion coefficients \cite{Mojarad2009,Orlov2012}. Here, $n$ is the multipole order and $m$ the azimuthal number, entering the multipoles as an exponential factor $e^{\imath m \phi}$. 

The decomposition of the scattered field into VSHs also enables one to represent the full scattering information of the nanoprism by its scattering matrix $\mathrm{\hat{T}}$ \cite{Tsang}. The electric field scattered off the nanoprism, $\textbf{E}_\text{sca}$, can thus written as:
\begin{align}
\textbf{E}_\text{sca}(\textbf{r}) = \mathrm{\hat{T}} \cdot \textbf{E}_\text{in}(\textbf{r}),
\end{align}
where $\textbf{E}_\text{in}$ represents the incident field expanded in VSHs. The scattering response of the nanoprism to different incident field structures and therefore VSHs can be calculated via boundary integrals \cite{Tsang}, where the shape of the triangular nanoprism can be analytically expressed by a spherical product of two superformulas \cite{Gielis2003}. These superformulas correspond to simple geometrical equations, enabling us to calculate the elements of the T-matrix by numerically integrating the analytical expressions. Looking at the T-matrix of the thus described particle, only the lowest-order entries show significant values for sub-wavelength nanoparticles. To simplify the interpretation of the possible multipolar interactions by the calculated T-matrix, it is diagonalized and thus decomposed into its Eigenvectors, representing the optical Eigenmodes of the nanoparticle.

\begin{figure*}[hbtp]%
\includegraphics[width=0.7\textwidth]{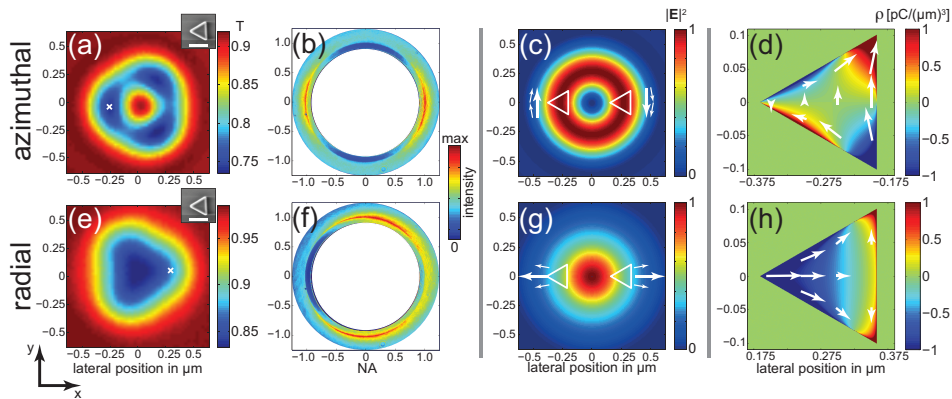}%
\caption{Experimental 2D-scan image of a single gold nanoprism excited by an (a) azimuthally and (e) radially polarized tightly focused beam. The insets show the orientation of the particle relative to the beam (scale bar = $\SI{200}{nm}$). (b) and (f) show the angular far-field emission of the particle at the marked point in (a) and (e), exhibiting mainly dipolar behavior. The actual size of the nanoprism with respect to the focal field is sketched in (c) and (g), with the corresponding overlapping local field sketched with white arrows. The numerically calculated charge density distributions on the particle for the positions indicated in (a) and (e) shown in (d) and (h) illustrate the quadrupolar contribution to the excited Eigenmode in the near-field. Here, the white arrows represent the current density distribution projected onto the transverse plane.}%
\label{fig:result}%
\end{figure*}

Applying this scheme to a gold nanoprism with the same dimensions as the one investigated in the experiment ($\SI{200}{nm}$ base-length, $\SI{50}{nm}$ thickness, refractive index following Johnson and Christy \cite{JohnsonChristy1972}, incident wavelength of $\lambda=\SI{630}{nm}$) results in two dominant degenerate Eigenmodes. They correspond to the lowest-order modes already described in other theoretical and experimental studies of similar systems \cite{Chuntonov2011}. Due to their degeneracy, there exists a subspace of Eigenmodes, in which one is free to choose different Eigenmode combinations spanning this subspace. This is used to define the electric dipole of one Eigenmode to be oriented along the x-axis, while the one of the other Eigenmode is chosen to be oriented along the y-axis (see Fig.\,\ref{fig:result}). The obtained two orthogonal Eigenmodes contain mainly four electric multipoles of order $\textbf{N}_{\pm 1,1}$ and $\textbf{N}_{\pm 2,2}$, with the amplitude of the transverse electric quadrupole amounting to only two percent of the amplitude of the electric dipole at the chosen wavelength ($|a_{\pm 2,2}|/|a_{\pm 1,1}| = 0.024$). This quadrupolar contribution in the Eigenmodes is responsible for the sensitivity to the local gradient of the electric field, i.e. its spatial distribution. In addition, the next Eigenmode contains a (nearly) pure electric z-dipole ($\textbf{N}_{0,1}$). Further higher-order Eigenmodes are only weakly excited for the chosen nanoprism parameters, and thus do not show up in our experimental scenario. The main challenge of all-optically probing the content of the described Eigenmodes of a single metallic nanoprism is thus to achieve a tailored excitation scheme of the nanoprism, allowing the resulting low-order multipolar distribution to be uniquely discriminated from the scanning map measured in the far-field.

Providing the nanoprism with a field distribution coinciding with one of the Eigenmodes would lead to an optimum excitation of the chosen mode, but has the downside of not exciting the other significant Eigenmode. Furthermore, one would have to rely on a priori information about the mode content.
Hence, we choose here a more flexible and versatile excitation field allowing for the realization of multiple excitation schemes just depending on the relative position of the sub-wavelength nanoprism within the field distribution. For this purpose, we utilize azimuthally and radially polarized light beams \cite{Quabis2000,Youngworth2000,Dorn2003}, which are tightly focused to create a diffraction-limited focal spot (see Fig.\,\ref{fig:setup}(a) and (b)). Such focal field distributions have the advantage of providing dominantly lower-order magnetic and electric multipole excitation for a sub-wavelength particle placed  on-axis \cite{Mojarad2009}. Additionally, scanning the particle through the focal field distribution leads to a position dependent coupling to different multipoles via the translation theorem for VSHs \cite{Cruzan1962}. This can be most intuitively understood by looking at the varying local electric field and its local gradients. Due to the axial symmetry of the field distribution, all possible excitations and low-order multipolar combinations are thus achieved within a single scan. As a consequence of the axial symmetry of the focal field, each excited individual multipole $\textbf{N}_{mn}$ or $\textbf{M}_{mn}$ would show an axially symmetric scanning response. Because of their different azimuthal phase dependence $e^{\imath m \phi}$, a sum of multipoles with different $m$ leads to a spatially varying interference signal of the scattered light. As a striking consequence, the distribution of a single scan image of the total transmitted light can be seen as a direct evidence of the low-order multipolar constituents of the excited fundamental Eigenmodes.

The experimental setup (similar to the one introduced	 in \cite{Banzer2010}) for implementing this multipole decomposition of low-order Eigenmodes is shown in Fig.\,\ref{fig:setup}(c). The initial beam is a linearly polarized Gaussian beam at a wavelength of $\lambda = \SI{630}{nm}$. This beam is impinging on a half-wave plate and a liquid crystal based polarization converter to convert the Gaussian mode into the desired azimuthally or radially polarized doughnut mode. Subsequently, the emerging beam is mode-cleaned by spatial Fourier-filtering, resulting in a $99\%$ overlap of the beam profile with the desired mode. The beam is then guided top-down onto the entrance pupil of a high numerical aperture (NA) microscope objective (NA=0.9) and tightly focused onto the sample.  The sample consists of single crystalline gold nanoprisms on a glass-substrate with a mean spacing of $\SI{5}{\micro\metre}$ (see inset in Fig.\,\ref{fig:setup}(c)). As already mentioned above, they are supporting low-order multipolar modes at the chosen wavelength. Positioning the sample in the focal plane of the microscope objective is achieved by means of a 3D piezo table, allowing for a position accuracy down to a few nanometer. The forward scattered and transmitted light is then collected by an oil-immersion microscope objective (NA=1.3). The back focal plane of the collecting microscope objective is imaged onto a camera to allow for an angularly resolved detection of the transmitted light.

For each position of the nanoparticle relative to the beam, the power of the total transmitted light is measured by integrating over the full back focal plane of the collecting objective (similar to Ref. \onlinecite{Bauer2014}). Thus, each entry of the resulting 2D-scan map is related to the extinction cross section of the nanoparticle at a certain position in the focal plane, measuring the interference signal between the forward scattered and transmitted light. Fig.\,\ref{fig:result}(a) and (e) show the experimental results of the interaction of the gold nanoprism with tightly focused azimuthally and radially polarized light. An SEM image of the nanoparticle with its orientation relative to the field distribution is depicted in the insets of the figures.

The most noticeable feature in the 2D-scan images is the already mentioned trifold symmetry of the particle response overlaid with the geometry of the corresponding focused beams. This trifold symmetry can be seen most clearly when exciting with an azimuthally polarized input field.  Furthermore, the angular spectrum of the  excited particle mode in the back focal plane of the collection objective (scattering pattern of the excited particle) is depicted in Fig.\,\ref{fig:result}(b) and (f), with the nanoprism located at a position where maximum scattering is observed. The angular range shown in these figures was chosen such, that only the scattering signal can be studied (the incoming beam is restricted to a solid angle below an NA of 0.9). The scattering of the particle shows an almost perfect dipolar pattern for the case of azimuthally polarized light. In addition, interference of the excited longitudinal and transverse electric dipoles oscillating $\pi/2$ out-of-phase is observed for radially polarized excitation, leading to the directive emission pattern in Fig.\,\ref{fig:result}(f) \cite{Neugebauer2014}. A significant and discernable contribution of the quadrupole content alters this symmetry of the angular spectrum. Depending on the respective quadrupole, the maxima of the angular scattering either show an asymmetry in their intensity or are angularly shifted towards each other. In our case, this contribution is hardly visible in a single far-field image (as opposed to a scanned image) due to the weak quadrupolar scattering response measured in the far-field. Nevertheless, when numerically calculating the electric charge and surface current density distribution for the same position of the particle relative to the beam via finite-difference time-domain analysis (FDTD Solutions, Lumerical Inc.), the higher-order mode contribution is clearly visible also in the near-field (see Fig.\,\ref{fig:result}(d) and (h)). 

The ability of the chosen scan approach for identifying even the weak multipolar contributions to the excited Eigenmode for different positions in a cylindrical symmetric focal field demonstrates the high mode sensitivity. The symmetry of the field allows for the excitation of the same Eigenmode for a given nanoprism orientation placed at certain positions in the focal field distribution on opposite sides relative to the optical axis, but with the overlap of the Eigenmode and the local field distribution, and thus the strength of excitation being different for both positions (see Fig.\,\ref{fig:result}(c) and (g)).

\begin{figure}[tbp]
\includegraphics[width=0.8\columnwidth]{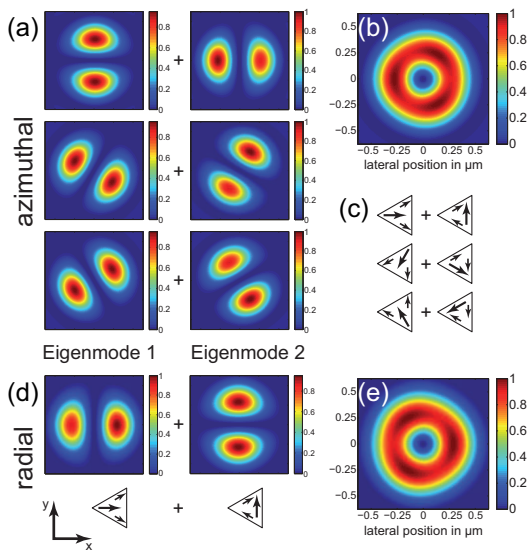}%
\caption{Theoretically calculated overlap of the first two Eigenmodes with a (a) azimuthally and (d) radially polarized tightly focused field distribution. The sum of the two orthogonal modes in (b) and (e) results in a trilobe scan image also observed experimentally. The same pattern can also be generated by the same Eigenmodes rotated by multiples of $\phi=2\pi/3$, as shown in (a). The sketches represent the induced current structure of the respective Eigenmode, with its specific orientation shown in (c).  All overlap distributions are normalized to the maximum of the sum of both Eigenmodes. }%
\label{fig:theory}%
\end{figure}

To further verify the interpretation of the trilobe intensity structure as direct evidence of the quadrupolar mode content in the dominant Eigenmodes of the nanoprism, we calculate the overlap of the modes with the azimuthally and radially polarized focal field distributions at each point in the focal plane (see Fig.\,\ref{fig:theory}(a) and (d), where the corresponding Eigenmode is sketched under the calculated distributions). This is achieved by determining the overlap integral of the theoretically calculated Eigenvectors of the T-matrix of the nanoprism with the field distribution in the focal plane, expanded into VSHs. The overlap integrals show that the scan image is mainly dominated by the different excitation strength of the Eigenmodes on opposite sides relative to the beam axis. The modes show a stronger excitation on the side that matches not only the local field direction but also its gradient, i.e. the quadrupole of the respective mode. While the amplitudes of the expansion coefficients of the beam are exactly the same at both positions ($|a^{'}_{\pm 2,2}|/|a^{'}_{\pm 1,1}| = 0.49$), the different phase of the dipolar contribution manifested by the oppositely pointing local field leads to a matching relative phase between the Eigenmode and the beam on one side of the optical axis. The other side offers a field with the relative phase exactly shifted by $\pi$ and therefore does not fit the mode structure, leading to a weaker excitation of the Eigenmode.  The sum of the response to both modes leads then to the trifold structure also seen in the experimental 2D scans (see Fig.\,\ref{fig:theory}(b) and (e)). This trifold structure can -- due to its symmetry -- also be generated by the same transverse Eigenmodes rotated by $\phi=2\pi/3$, as shown exemplarily for the azimuthally polarized light in Fig.\,\ref{fig:theory}(a). The orientation of the Eigenmodes for these cases is sketched in Fig.\,\ref{fig:theory}(c).  For the case of an excitation with tightly focused radially polarized light, the additional electric dipole Eigenmode (oscillating along the z-axis) would contribute to the scattering signal, leading to the measured 2D-scan image with non-zero scattering for the particle sitting on-axis (see Fig.\,\ref{fig:result}(e)). With this decomposition, the trifold scattering distribution is confirmed to be governed by the presence of the transverse electric quadrupole components with $m=\pm 2$  and electric dipole components with $m=\mp 1$.

In conclusion, we have experimentally demonstrated how to investigate the multipolar constituents of the Eigenmodes of single metallic nanoprisms by means of cylindrical vector beams. By choosing tightly focused azimuthally and radially polarized light beams providing axially symmetric focal field distributions, we were able to encode the multipolar information of the Eigenmodes of the nanoprism in the 2D-scanning map. The described method allows for a fast and precise characterization of the low-order Eigenmodes of arbitrary single nanoparticles.

\bibliography{far_field_scattering}

\end{document}